\pdfoutput=1

\documentclass[12pt,a4paper]{article}

\usepackage{ifthen} 
\newboolean{pdflatex}
\setboolean{pdflatex}{true} 

\newboolean{articletitles}
\setboolean{articletitles}{true} 

\newboolean{uprightparticles}
\setboolean{uprightparticles}{false} 

\newboolean{inbibliography}
\setboolean{inbibliography}{false} 

\def\paperauthors{LHCb collaboration} 
\def\paperasciititle{Observation of the doubly charmed baryon Xicc++} 
\def\papertitle{Observation of the doubly charmed baryon $\Xiccpp$} 
\def\paperkeywords{{High Energy Physics}, {LHCb}} 
\def\papercopyright{CERN on behalf of the LHCb collaboration}
\def\paperlicence{CC-BY-4.0}
\def\paperlicenceurl{https://creativecommons.org/licenses/by/4.0/}

\usepackage[top=1in, bottom=1.25in, left=1in, right=1in]{geometry}

\usepackage{microtype}
\usepackage{lineno}  
\usepackage{xspace} 
\usepackage{caption} 

\usepackage{graphicx}  
\usepackage{color}
\usepackage{colortbl}
\graphicspath{{./figs/}} 

\usepackage{amsmath} 
\usepackage{amssymb}
\usepackage{amsfonts}
\usepackage{upgreek} 

\newcommand*\patchAmsMathEnvironmentForLineno[1]{%
\expandafter\let\csname old#1\expandafter\endcsname\csname #1\endcsname
\expandafter\let\csname oldend#1\expandafter\endcsname\csname
end#1\endcsname
 \renewenvironment{#1}%
   {\linenomath\csname old#1\endcsname}%
   {\csname oldend#1\endcsname\endlinenomath}%
}
\newcommand*\patchBothAmsMathEnvironmentsForLineno[1]{%
  \patchAmsMathEnvironmentForLineno{#1}%
  \patchAmsMathEnvironmentForLineno{#1*}%
}
\AtBeginDocument{%
\patchBothAmsMathEnvironmentsForLineno{equation}%
\patchBothAmsMathEnvironmentsForLineno{align}%
\patchBothAmsMathEnvironmentsForLineno{flalign}%
\patchBothAmsMathEnvironmentsForLineno{alignat}%
\patchBothAmsMathEnvironmentsForLineno{gather}%
\patchBothAmsMathEnvironmentsForLineno{multline}%
\patchBothAmsMathEnvironmentsForLineno{eqnarray}%
}

\usepackage{hyperxmp}

\usepackage[pdftex,
            pdfauthor={\paperauthors},
            pdftitle={\paperasciititle},
            pdfkeywords={\paperkeywords},
            pdfcopyright={Copyright (C) \papercopyright},
            pdflicenseurl={\paperlicenceurl}]{hyperref}

\usepackage[all]{hypcap} 

\usepackage{xspace} 
\usepackage{upgreek}

\def\lhcb {\mbox{LHCb}\xspace}

\def\babar  {\mbox{BaBar}\xspace}
\def\belle  {\mbox{Belle}\xspace}

\def\MagUp {\mbox{\em Mag\kern -0.05em Up}\xspace}

\ifthenelse{\boolean{uprightparticles}}%
{

 \def\Pmu         {\ensuremath{\upmu}\xspace}

 \def\Ppi         {\ensuremath{\uppi}\xspace}                 
                  
 \def\Prho        {\ensuremath{\uprho}\xspace}

 \def\Ppsi        {\ensuremath{\uppsi}\xspace}

 \def\PDelta      {\ensuremath{\Delta}\xspace}                 
 \def\PXi      {\ensuremath{\Xi}\xspace}                 
 \def\PLambda      {\ensuremath{\Lambda}\xspace}                 
 \def\PSigma      {\ensuremath{\Sigma}\xspace}                 
 \def\POmega      {\ensuremath{\Omega}\xspace}                 
 \def\PUpsilon      {\ensuremath{\Upsilon}\xspace}                 
 
 \def\PB      {\ensuremath{\mathrm{B}}\xspace}                 
                  
 \def\PD      {\ensuremath{\mathrm{D}}\xspace}

 \def\PJ      {\ensuremath{\mathrm{J}}\xspace}                 
 \def\PK      {\ensuremath{\mathrm{K}}\xspace}

 \def\Pb      {\ensuremath{\mathrm{b}}\xspace}                 
 \def\Pc      {\ensuremath{\mathrm{c}}\xspace}                 
 \def\Pd      {\ensuremath{\mathrm{d}}\xspace}

 \def\Pi      {\ensuremath{\mathrm{i}}\xspace}

 \def\Pp      {\ensuremath{\mathrm{p}}\xspace}

 \def\Ps      {\ensuremath{\mathrm{s}}\xspace}                 
                  
 \def\Pu      {\ensuremath{\mathrm{u}}\xspace}

}
{

 \def\Pmu         {\ensuremath{\mu}\xspace}

 \def\Ppi         {\ensuremath{\pi}\xspace}                 
                  
 \def\Prho        {\ensuremath{\rho}\xspace}

 \def\Ppsi        {\ensuremath{\psi}\xspace}                 
                  
 \mathchardef\PDelta="7101
 \mathchardef\PXi="7104
 \mathchardef\PLambda="7103
 \mathchardef\PSigma="7106
 \mathchardef\POmega="710A
 \mathchardef\PUpsilon="7107
                  
 \def\PB      {\ensuremath{B}\xspace}                 
                  
 \def\PD      {\ensuremath{D}\xspace}

 \def\PJ      {\ensuremath{J}\xspace}                 
 \def\PK      {\ensuremath{K}\xspace}

 \def\Pb      {\ensuremath{b}\xspace}                 
 \def\Pc      {\ensuremath{c}\xspace}                 
 \def\Pd      {\ensuremath{d}\xspace}

 \def\Pi      {\ensuremath{i}\xspace}

 \def\Pp      {\ensuremath{p}\xspace}

 \def\Ps      {\ensuremath{s}\xspace}                 
                  
 \def\Pu      {\ensuremath{u}\xspace}

}

\makeatletter
\ifcase \@ptsize \relax
  \newcommand{\miniscule}{\@setfontsize\miniscule{4}{5}}
\or
  \newcommand{\miniscule}{\@setfontsize\miniscule{5}{6}}
\or
  \newcommand{\miniscule}{\@setfontsize\miniscule{5}{6}}
\fi
\makeatother

\DeclareRobustCommand{\optbar}[1]{\shortstack{{\miniscule (\rule[.5ex]{1.25em}{.18mm})}
  \\ [-.7ex] $#1$}}

\def\mumu       {{\ensuremath{\Pmu^+\Pmu^-}}\xspace}

\def\uquark    {{\ensuremath{\Pu}}\xspace}

\def\dquark    {{\ensuremath{\Pd}}\xspace}

\def\squark    {{\ensuremath{\Ps}}\xspace}

\def\cquark    {{\ensuremath{\Pc}}\xspace}

\def\bquark    {{\ensuremath{\Pb}}\xspace}

\def\pion   {{\ensuremath{\Ppi}}\xspace}

\def\pip    {{\ensuremath{\pion^+}}\xspace}
\def\pim    {{\ensuremath{\pion^-}}\xspace}
\def\pipm   {{\ensuremath{\pion^\pm}}\xspace}

\def\rhomeson {{\ensuremath{\Prho}}\xspace}
\def\rhoz     {{\ensuremath{\rhomeson^0}}\xspace}

\def\kaon    {{\ensuremath{\PK}}\xspace}
  \def\Kbar    {{\kern 0.2em\overline{\kern -0.2em \PK}{}}\xspace}

\def\KorKbar    {\kern 0.18em\optbar{\kern -0.18em K}{}\xspace}

\def\Kp      {{\ensuremath{\kaon^+}}\xspace}
\def\Km      {{\ensuremath{\kaon^-}}\xspace}

\def\Kstarz  {{\ensuremath{\kaon^{*0}}}\xspace}

  \def\Dbar    {{\kern 0.2em\overline{\kern -0.2em \PD}{}}\xspace}
\def\D       {{\ensuremath{\PD}}\xspace}

\def\DorDbar    {\kern 0.18em\optbar{\kern -0.18em D}{}\xspace}

\def\Dp      {{\ensuremath{\D^+}}\xspace}

\def\Ds      {{\ensuremath{\D^+_\squark}}\xspace}

\def\B       {{\ensuremath{\PB}}\xspace}
\def\Bbar    {{\ensuremath{\kern 0.18em\overline{\kern -0.18em \PB}{}}}\xspace}

\def\BorBbar    {\kern 0.18em\optbar{\kern -0.18em B}{}\xspace}

\def\Bu      {{\ensuremath{\B^+}}\xspace}

\def\Bp      {{\ensuremath{\Bu}}\xspace}

\def\jpsi     {{\ensuremath{{\PJ\mskip -3mu/\mskip -2mu\Ppsi\mskip 2mu}}}\xspace}

  \def\Y#1S{\ensuremath{\PUpsilon{(#1S)}}\xspace}

\def\proton      {{\ensuremath{\Pp}}\xspace}

\def\Xires       {{\ensuremath{\PXi}}\xspace}

\def\Lz          {{\ensuremath{\PLambda}}\xspace}
\def\Lbar        {{\ensuremath{\kern 0.1em\overline{\kern -0.1em\PLambda}}}\xspace}
\def\LorLbar    {\kern 0.18em\optbar{\kern -0.18em \PLambda}{}\xspace}

\def\Sigmares    {{\ensuremath{\PSigma}}\xspace}

\def\Omegares    {{\ensuremath{\POmega}}\xspace}

\def\Lc      {{\ensuremath{\Lz^+_\cquark}}\xspace}

\def\to                 {\ensuremath{\rightarrow}\xspace}

\def\AT#1     {\ensuremath{A_{\mathrm{T}}^{#1}}\xspace}           

\def\C#1      {\ensuremath{\mathcal{C}_{#1}}\xspace}                       
\def\Cp#1     {\ensuremath{\mathcal{C}_{#1}^{'}}\xspace}                    
\def\Ceff#1   {\ensuremath{\mathcal{C}_{#1}^{\mathrm{(eff)}}}\xspace}        
\def\Cpeff#1  {\ensuremath{\mathcal{C}_{#1}^{'\mathrm{(eff)}}}\xspace}       
\def\Ope#1    {\ensuremath{\mathcal{O}_{#1}}\xspace}                       
\def\Opep#1   {\ensuremath{\mathcal{O}_{#1}^{'}}\xspace}                    

\newcommand{\tev}{\ifthenelse{\boolean{inbibliography}}{\ensuremath{~T\kern -0.05em eV}}{\ensuremath{\mathrm{\,Te\kern -0.1em V}}}\xspace}
\newcommand{\gev}{\ensuremath{\mathrm{\,Ge\kern -0.1em V}}\xspace}
\newcommand{\mev}{\ensuremath{\mathrm{\,Me\kern -0.1em V}}\xspace}
\newcommand{\kev}{\ensuremath{\mathrm{\,ke\kern -0.1em V}}\xspace}
\newcommand{\ev}{\ensuremath{\mathrm{\,e\kern -0.1em V}}\xspace}
\newcommand{\gevc}{\ensuremath{{\mathrm{\,Ge\kern -0.1em V\!/}c}}\xspace}
\newcommand{\mevc}{\ensuremath{{\mathrm{\,Me\kern -0.1em V\!/}c}}\xspace}
\newcommand{\gevcc}{\ensuremath{{\mathrm{\,Ge\kern -0.1em V\!/}c^2}}\xspace}
\newcommand{\gevgevcccc}{\ensuremath{{\mathrm{\,Ge\kern -0.1em V^2\!/}c^4}}\xspace}
\newcommand{\mevcc}{\ensuremath{{\mathrm{\,Me\kern -0.1em V\!/}c^2}}\xspace}

\def\invfb   {\ensuremath{\mbox{\,fb}^{-1}}\xspace}

\def\fs   {\ensuremath{\mathrm{ \,fs}}\xspace}

\newcommand{\stat}{\ensuremath{\mathrm{\,(stat)}}\xspace}
\newcommand{\syst}{\ensuremath{\mathrm{\,(syst)}}\xspace}

\newcommand{\chisq}{\ensuremath{\chi^2}\xspace}

\newcommand{\chisqip}{\ensuremath{\chi^2_{\text{IP}}}\xspace}

\def\gsim{{~\raise.15em\hbox{$>$}\kern-.85em
          \lower.35em\hbox{$\sim$}~}\xspace}
\def\lsim{{~\raise.15em\hbox{$<$}\kern-.85em
          \lower.35em\hbox{$\sim$}~}\xspace}

\def\pt         {\mbox{$p_{\mathrm{ T}}$}\xspace}

\def\mrad{\ensuremath{\mathrm{ \,mrad}}\xspace}

\def\tell1  {TELL1\xspace}
\def\ukl1   {UKL1\xspace}

\newcommand{\eg}{\mbox{\itshape e.g.}\xspace}
\newcommand{\ie}{\mbox{\itshape i.e.}\xspace}

\usepackage{cite} 
\usepackage{mciteplus}

\usepackage{xspace} 
\usepackage{upgreek}

\def\DeltaOrDeltabar  {\kern 0.18em\optbar{\kern -0.18em \PDelta}{}\xspace}
\def\XiOrXibar        {\kern 0.18em\optbar{\kern -0.18em \PXi}{}\xspace}
\def\SigmaOrSigmabar  {\kern 0.18em\optbar{\kern -0.18em \PSigma}{}\xspace}
\def\OmegaOrOmegabar  {\kern 0.18em\optbar{\kern -0.18em \POmega}{}\xspace}

\def\Lcp          {{\ensuremath{\Lz^+_\cquark}}\xspace}

\def\Sigmacpp     {{\ensuremath{\Sigmares^{++}_\cquark}}\xspace}

\def\Sigmacz      {{\ensuremath{\Sigmares^{0}_\cquark}}\xspace}

\def\Xiccbare     {{\ensuremath{\Xires_{\cquark\cquark}}}\xspace}
\def\Xiccpp       {{\ensuremath{\Xires^{++}_{\cquark\cquark}}}\xspace}

\def\Xiccp        {{\ensuremath{\Xires^{+}_{\cquark\cquark}}}\xspace}

\def\Omegaccp     {{\ensuremath{\Omegares^{+}_{\cquark\cquark}}}\xspace}

\newcommand{\TeVnosp}{\ifthenelse{\boolean{inbibliography}}{\ensuremath{~T\kern -0.05em eV}}{\ensuremath{\mathrm{Te\kern -0.1em V}}}}
\newcommand{\GeVnosp}{\ensuremath{\mathrm{Ge\kern -0.1em V}}}
\newcommand{\MeVnosp}{\ensuremath{\mathrm{Me\kern -0.1em V}}}
\newcommand{\keVnosp}{\ensuremath{\mathrm{ke\kern -0.1em V}}}
\newcommand{\eVnosp}{\ensuremath{\mathrm{e\kern -0.1em V}}}
\newcommand{\GeVcnosp}{\ensuremath{{\mathrm{Ge\kern -0.1em V\!/}c}}}
\newcommand{\MeVcnosp}{\ensuremath{{\mathrm{Me\kern -0.1em V\!/}c}}}
\newcommand{\GeVccnosp}{\ensuremath{{\mathrm{Ge\kern -0.1em V\!/}c^2}}}
\newcommand{\MeVccnosp}{\ensuremath{{\mathrm{Me\kern -0.1em V\!/}c^2}}}
\newcommand{\GeVGeVccccnosp}{\ensuremath{{\mathrm{Ge\kern -0.1em V^2\!/}c^4}}}

\def\genxicc      {\mbox{\textsc{GENXICC}}\xspace}

\def\XiccppDecay  {{\ensuremath{\Xiccpp\to\Lcp\Km\pip\pip}}\xspace}

\def\genxicc      {\mbox{\textsc{Genxicc}}\xspace}

\newcommand{\nogapmevcc}{\ensuremath{{\mathrm{Me\kern -0.1em V\!/}c^2}}\xspace}

\def\mcand {{\ensuremath{m_{\mathrm{cand}}}}\xspace}

\def\XiccDeltaMass {{\ensuremath{1334.94 \pm 0.72 \stat \pm 0.27 \syst \mevcc}}\xspace}
\def\XiccmassWithLcError {{\ensuremath{3621.40 \pm 0.72 \stat \pm 0.27 \syst \pm 0.14 \, (\Lc) \mevcc}}\xspace}

\def\xiccplotwidth{0.48\textwidth}
\def\xiccfeynwidth{0.7\textwidth}
\def\xiccfirstplot{left}
\def\xiccsecondplot{right}

\begin{document}

\renewcommand{\thefootnote}{\fnsymbol{footnote}}
\setcounter{footnote}{1}


\begin{titlepage}
\pagenumbering{roman}

\vspace*{-1.5cm}
\centerline{\large EUROPEAN ORGANIZATION FOR NUCLEAR RESEARCH (CERN)}
\vspace*{1.5cm}
\noindent
\begin{tabular*}{\linewidth}{lc@{\extracolsep{\fill}}r@{\extracolsep{0pt}}}
\ifthenelse{\boolean{pdflatex}}
{\vspace*{-2.7cm}\mbox{\!\!\!\includegraphics[width=.14\textwidth]{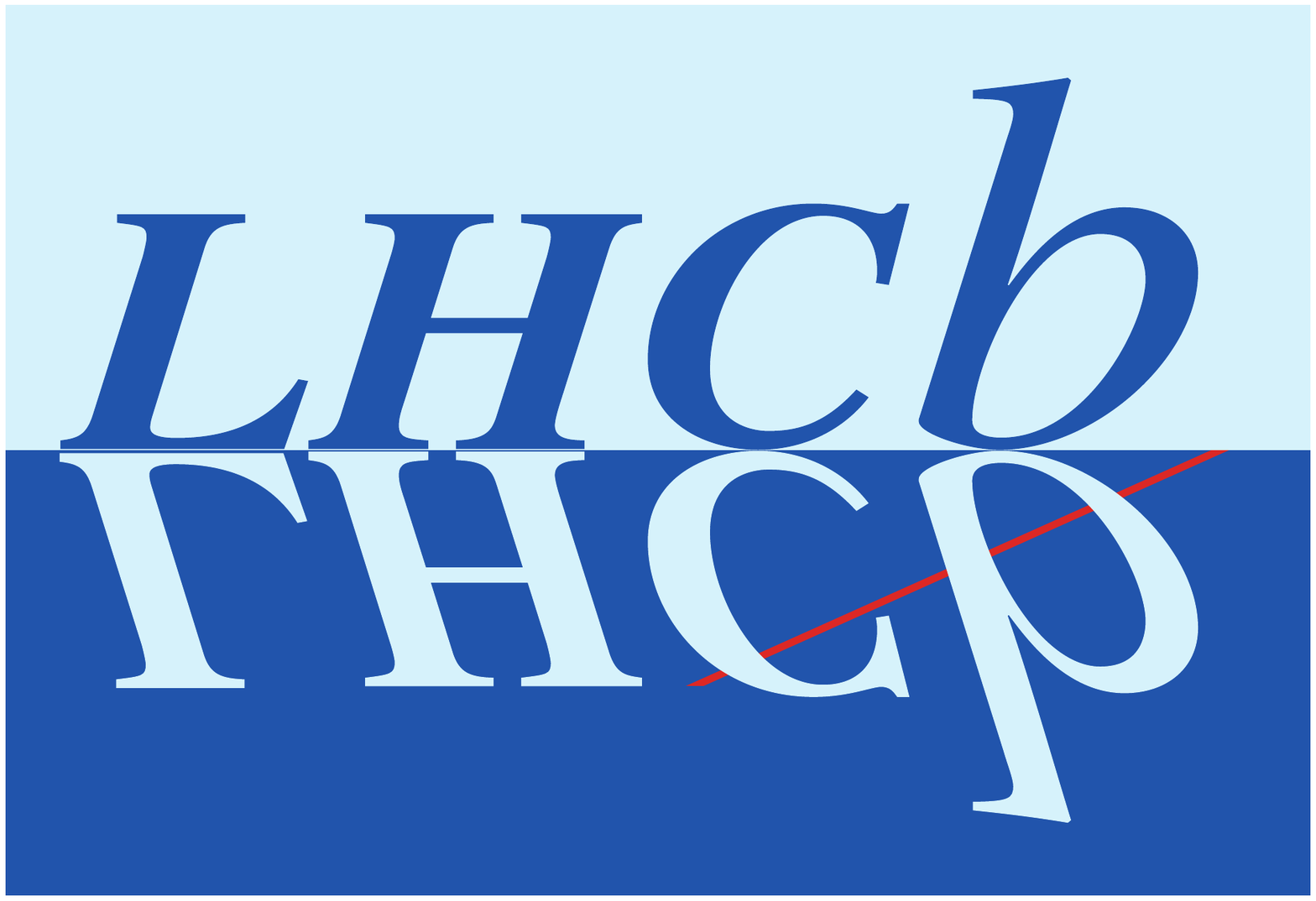}} & &}%
{\vspace*{-1.2cm}\mbox{\!\!\!\includegraphics[width=.12\textwidth]{lhcb-logo.eps}} & &}%
\\
 & & CERN-EP-2017-156 \\  
 & & LHCb-PAPER-2017-018 \\  
 & & 12 September 2017 \\ 
\end{tabular*}

\vspace*{4.0cm}

{\normalfont\bfseries\boldmath\huge
\begin{center}
  \papertitle 
\end{center}
}

\vspace*{2.0cm}

\begin{center}
\paperauthors\footnote{Authors are listed at the end of this paper.}
\end{center}

\vspace{\fill}

\begin{abstract}
  \noindent
  A highly significant structure is observed in the $\Lcp\Km\pip\pip$
  mass spectrum, where the \Lcp baryon is reconstructed
  in the decay mode $\proton\Km\pip$.
  The structure is consistent with originating from a weakly decaying particle,
  identified as the doubly charmed baryon \Xiccpp.
  The difference between the masses of the $\Xiccpp$ and $\Lcp$ states is measured to be
  $\XiccDeltaMass$, and the $\Xiccpp$ mass is then
  determined to be $\XiccmassWithLcError$,
  where the last uncertainty is due to the limited knowledge of the $\Lcp$ mass.
  The state is observed in a sample of proton-proton collision data collected
  by the LHCb experiment at a center-of-mass energy of 13\tev, corresponding to an integrated luminosity of $1.7$\invfb,
  and confirmed in an additional sample
  of data collected at 8\tev.

\end{abstract}

\vspace*{2.0cm}

\begin{center}
  Published in Phys.~Rev.~Lett. 119 (2017) 112001
\end{center}

\vspace{\fill}

{\footnotesize 
\centerline{\copyright~\papercopyright, license \href{\paperlicenceurl}{\paperlicence}.}}
\vspace*{2mm}

\end{titlepage}


\newpage
\setcounter{page}{2}
\mbox{~}
%
%
%
%

\cleardoublepage

\renewcommand{\thefootnote}{\arabic{footnote}}
\setcounter{footnote}{0}


\pagestyle{plain} 
\setcounter{page}{1}
\pagenumbering{arabic}

The quark model~\cite{GellMann:1964nj,Zweig:1981pd,Zweig:1964jf} predicts the existence of multiplets of baryon and meson states.
Those states composed of the lightest four quarks (\uquark,\,\dquark,\,\squark,\,\cquark) form $SU(4)$ multiplets~\cite{DeRujula:1975ge}. 
Numerous states with charm quantum number $C=0$ or $C=1$ have been discovered, including all of the expected $q \bar{q}$ and $qqq$ ground states~\cite{PDG2017}.
Three weakly decaying $qqq$ states with $C=2$ are expected:
one isospin doublet
($\Xiccpp = \cquark\cquark\uquark$ and $\Xiccp = \cquark\cquark\dquark$)
and one  isospin singlet ($\Omegaccp=\cquark\cquark\squark$),
each with spin-parity $J^P = 1/2^+$.
The properties of these baryons have been calculated with a variety of theoretical models. In most cases, the masses of
the \Xiccbare states are predicted to lie in the range 3500 to 3700\mevcc~\cite{Gershtein:1998un,Gershtein:1998sx,Itoh:2000um,Gershtein:2000nx,Anikeev:2001rk,Kiselev:2001fw,Ebert:2002ig,He:2004px,Chang:2006eu,Roberts:2007ni,Valcarce:2008dr,Zhang:2008rt,wang2010analysis,Karliner:2014gca,Wei:2015gsa,Sun:2016wzh,Alexandrou:2017xwd,Kerbikov:1987vx,Fleck:1989mb,Chernyshev:1995gj,Aliev:2012ru,Sun:2014aya,Mathur:2002ce,Namekawa:2013vu,Brown:2014ena,Padmanath:2015jea,Bali:2015lka,Liu:2017xzo,*Liu:2017frj}.
The masses of the \Xiccpp and \Xiccp states are expected to differ by only a
few~\mevcc, due to approximate isospin
symmetry~\cite{Hwang:2008dj,Brodsky:2011zs,Karliner:2017gml}.
Most predictions for the lifetime of the \Xiccp baryon are in the range
50 to 250\fs, and the lifetime of the \Xiccpp baryon is expected to be
three to four times longer at 200 to 700\fs~\cite{guberina1999inclusive,Kiselev:1998sy,Anikeev:2001rk,Kiselev:2001fw,hsi2008lifetime,Karliner:2014gca,Berezhnoy:2016wix,Fleck:1989mb}.
While both are expected to be produced at hadron colliders~\cite{Berezhnoy:1994ba,Kolodziej:1995nv,Berezhnoy:1998aa},
the longer lifetime of the \Xiccpp baryon should make it significantly easier
to observe than the \Xiccp baryon in such experiments, due to the use of real-time (online) event-selection requirements designed to
reject backgrounds originating from the primary interaction point.

Experimentally, there is a longstanding puzzle in the \Xiccbare system.
Observations of the \Xiccp baryon
at a mass of $3519 \pm 2$\mevcc 
with signal yields of 15.9 events over $6.1 \pm 0.5$ background
in the final state $\Lcp \Km \pip$ 
($6.3\sigma$ significance), and 5.62 events over $1.38 \pm 0.13$ background
in the final state $\proton \Dp \Km$
($4.8\sigma$ significance) were reported by the SELEX collaboration~\cite{Mattson:2002vu,Ocherashvili:2004hi}.
Their results included a number of unexpected features,
notably a short lifetime and a large production rate relative to that of the singly charmed $\Lcp$ baryon. The lifetime
was stated to be shorter than $33\fs$ at the 90\% confidence level, and SELEX concluded that 20\% of all $\Lcp$ baryons observed by the experiment originated
from $\Xiccp$ decays, implying a relative \Xiccbare production rate several
orders of magnitude larger than theoretical expectations~\cite{Kiselev:2001fw}.
Searches from the FOCUS~\cite{Ratti:2003ez}, \babar~\cite{Aubert:2006qw}, and \belle~\cite{Chistov:2006zj} experiments did not find
evidence for a state with the properties reported by
SELEX,
and neither did a search at LHCb with data collected in
2011 corresponding to an integrated luminosity of $0.65$\invfb~\cite{LHCb-PAPER-2013-049}.
However, because the production environments at these experiments differ
from that of SELEX, which studied collisions of a hyperon beam
on fixed nuclear targets,
these null results do not exclude the original observations.

\begin{figure}
\begin{center}
\includegraphics[width=\xiccfeynwidth]{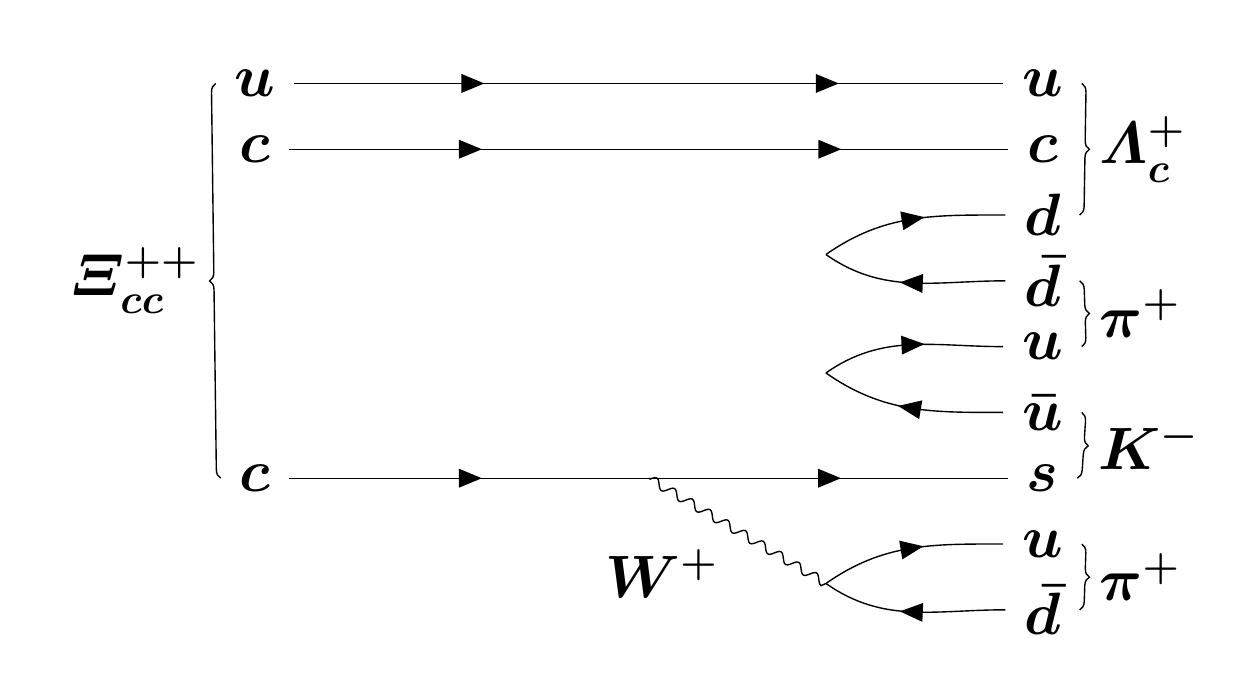}
\end{center}
\caption{
  Example Feynman diagram contributing to the decay \XiccppDecay.
}
\label{fig:FeynmanDiagram}
\end{figure}

This Letter presents the observation of the \Xiccpp baryon\footnote{ 
  Inclusion of charge-conjugate processes is implied throughout.} via the decay mode
$\Lcp\Km\pip\pip$ (Fig.~\ref{fig:FeynmanDiagram}),
which is expected to have a branching fraction of up to
10\%~\cite{Yu:2017zst}.
The \Lcp baryon is reconstructed in the final state $\proton\Km\pip$.
The data consist of $pp$ collisions collected by the LHCb
experiment at the Large Hadron Collider at CERN
with a center-of-mass energy of $13\tev$ taken in 2016, 
corresponding to an integrated luminosity of $1.7\invfb$.

The LHCb detector is a single-arm forward spectrometer covering the \mbox{pseudorapidity} range $2<\eta <5$, designed for the study of particles containing \bquark or \cquark quarks, and is described in detail in Refs.~\cite{Alves:2008zz,LHCb-DP-2014-002}.
The detector elements most
relevant to this analysis are a silicon-strip vertex detector surrounding the $pp$ interaction
region, a tracking system that provides a measurement of the momentum of charged
particles, and two ring-imaging Cherenkov detectors~\cite{LHCb-DP-2012-003} that are able to discriminate between
different species of charged hadrons.
The online event selection is performed by a trigger that consists of a hardware stage, which is based on information from the calorimeter and muon systems, followed by a software stage, which fully reconstructs the event~\cite{LHCb-DP-2012-004}.
The online reconstruction incorporates near-real-time alignment and calibration of the
detector~\cite{LHCb-PROC-2015-011}, which in turn allows the reconstruction of the \Xiccpp
decay to be performed entirely in the trigger software.

The reconstruction of \XiccppDecay decays proceeds as follows.
Candidate $\Lcp \to \proton \Km \pip$ decays are reconstructed
from three charged particles 
  that form a good-quality vertex
and
  that are inconsistent with originating from any $\proton \proton$ collision primary vertex (PV).
The PV of any single particle is defined to be
the PV with respect to which
the particle has the smallest impact parameter $\chi^2$ (\chisqip), which is
the difference in $\chi^2$ of the PV fit
with and without the particle in question.
The \Lcp vertex is required to be displaced from its PV
by a distance corresponding to a proper decay time greater than 150\fs.
The \Lcp candidate is then combined with three additional charged particles
  to form a \XiccppDecay candidate.
These additional particles must form a good-quality vertex with the \Lcp candidate, and
the \Lc decay vertex must be downstream of the \Xiccpp vertex.
Each of the six final-state particles is required to pass track-quality
  requirements,
  to have hadron-identification information consistent with the appropriate hypothesis
    (\proton, \kaon, or \pion),
  and to have transverse momentum $\pt > 500$\mevc.
To avoid duplicate tracks, the angle between each pair of final-state particles with the same charge is required to be larger than $0.5$\mrad.
  The $\Xiccpp$ candidate must have $\pt > 4$\gevc
and  must be consistent with originating from its PV.
%
The selection above includes criteria applied in the trigger software,
plus additional requirements
chosen based on simulated signal events and a control sample of data.
Simulated signal events are 
produced with the standard LHCb simulation software~\cite{Sjostrand:2007gs,Sjostrand:2006za,LHCb-PROC-2010-056,Lange:2001uf,Golonka:2005pn,Agostinelli:2002hh,*Allison:2006ve,LHCb-PROC-2011-006}
interfaced to
a dedicated generator, \genxicc~\cite{Chang:2007pp,Chang:2009va,Wang:2012vj}, for
\Xiccpp baryon production.
In the simulation,
the \Xiccpp mass and lifetime are assumed to be $3600$\mevcc and 333\fs.
The background control sample consists of wrong-sign (WS)
$\Lcp\Km\pip\pim$ combinations.

The background level is further reduced with a multivariate selector
based on the multilayer perceptron algorithm~\cite{Hocker:2007ht}.
The selector is trained with simulated signal events
and with the WS control sample of data to represent the background.
For both signal and background training samples, candidates
are required to pass the selection described above and
to fall within a signal search region defined as
$2270 < \mcand(\Lcp) < 2306$\mevcc and
$3300 < \mcand(\Xiccpp) < 3800$\mevcc,
where $\mcand(\Lcp)$ is the reconstructed mass of the \Lcp candidate,
$\mcand(\Xiccpp) \equiv m(\Lcp \Km \pip \pipm) - \mcand(\Lcp) + m_{\mathrm{PDG}}(\Lcp)$,
$m(\Lcp \Km \pip \pipm)$ is the reconstructed mass of the $\Lcp \Km \pip \pipm$
combination,
and 
$m_{\mathrm{PDG}}(\Lcp) = 2286.46 \pm 0.14$\mevcc 
is the known value of the \Lcp mass~\cite{PDG2017}.
The $\mcand(\Lcp)$ window corresponds
to approximately $\pm 3$ times the \Lcp mass resolution.
The use of $\mcand(\Xiccpp)$ rather than $m(\Lcp \Km \pip \pipm)$
cancels fluctuations in the reconstructed \Lc mass to first order,
and thereby improves the \Xiccpp mass resolution by approximately 40\%.

Based on studies with simulated events and control samples of data, ten
input variables that together provide good discrimination between signal and
background candidates
are used in the multivariate selector. They are as follows:
  the $\chi^2$ per degree of freedom of each of the \Lc vertex fit,
    the \Xiccpp vertex fit,
    and a kinematic refit~\cite{Hulsbergen:2005pu} of the \Xiccpp decay chain requiring it to originate from its PV;
  the smallest \pt of the three decay products of the \Lcp;
  the smallest \pt of the four decay products of the \Xiccpp;
  the scalar sum of the \pt of the four decay products of the \Xiccpp;
  the angle between the \Xiccpp momentum vector and the direction from the PV to the \Xiccpp decay vertex;
  the flight distance $\chisq$ between the PV and the $\Xiccpp$ decay vertex;
  the $\chisqip$ of the \Xiccpp with respect to its PV; and
  the smallest $\chisqip$ of the decay products of the \Xiccpp with respect to its PV.
Here,
the flight distance $\chi^2$ is defined as the $\chi^2$ of the hypothesis that the $\Xiccpp$ decay vertex coincides with its PV.
Candidates are retained for analysis only if their multivariate selector output values
exceed a threshold
chosen by maximizing the expected
value of the figure of merit 
$\varepsilon/(\frac{5}{2}+\sqrt{B})$~\cite{Punzi:2003bu},
where $\varepsilon$ is the estimated signal efficiency and
  $B$ is the estimated number of background candidates underneath the signal peak.
The quantity $B$ is computed with the WS control sample and,
purely for the purposes of this optimization, it is calculated
in a window centered at a mass of 3600\mevcc
and of halfwidth $12.5$\mevcc (corresponding to
approximately twice the expected resolution).
Its evaluation takes into account the difference in background rates between
the $\Lcp \Km \pip \pip$ signal mode and the WS sample, scaling the WS background
by the ratio seen in data in the sideband regions
$3200<\mcand(\Xiccpp)<3300$\mevcc and $3800 < \mcand(\Xiccpp) < 3900$\mevcc.
The performance of the multivariate selector is also tested for simulated signal
events under other lifetime hypotheses; while the signal efficiency increases with the lifetime,
it is found that the training obtained for 333\fs is close to optimal
(\ie gives comparable performance to a training optimized for the new lifetime hypothesis)
even for much shorter or longer lifetimes.

After the multivariate selection is applied, events
may still contain more than one \Xiccpp candidate in the signal search region.
Based on studies of simulation and the control data sample,
no peaking background arises due to multiple candidates
except for the special case in which the candidates are formed from the same six decay
products but two of the decay products are interchanged
(\eg, the \Km particle from the \Xiccpp decay and the \Km particle from the \Lcp decay).
In such instances, one of the candidates
is chosen at random to be retained and all others are discarded.
In the remaining events, the fraction that has more than one \Xiccpp candidate
in the range 3300--3800\mevcc is approximately 8\%.

\begin{figure}
  \begin{center}
    \includegraphics[width=\xiccplotwidth]{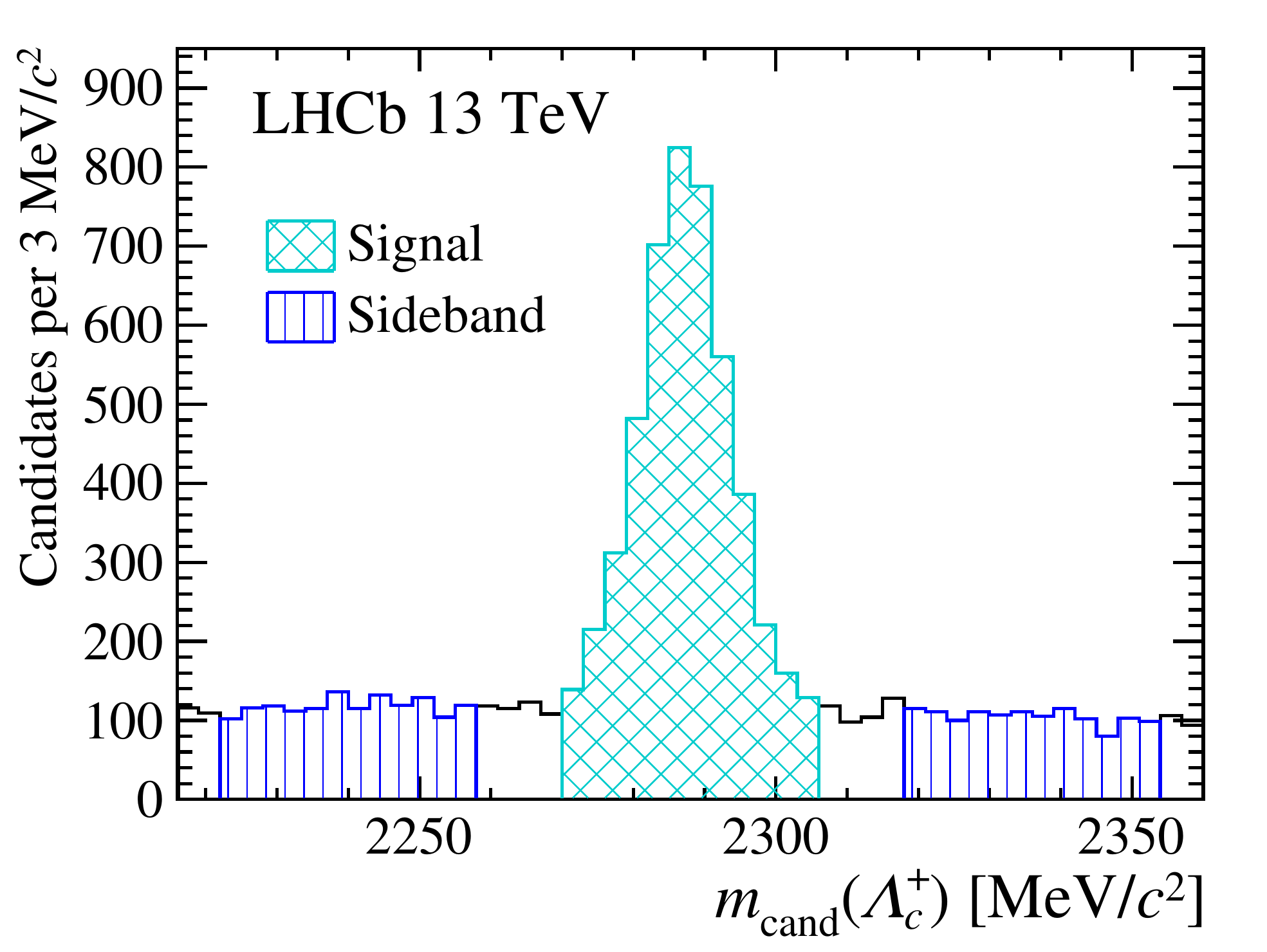}
    \includegraphics[width=\xiccplotwidth]{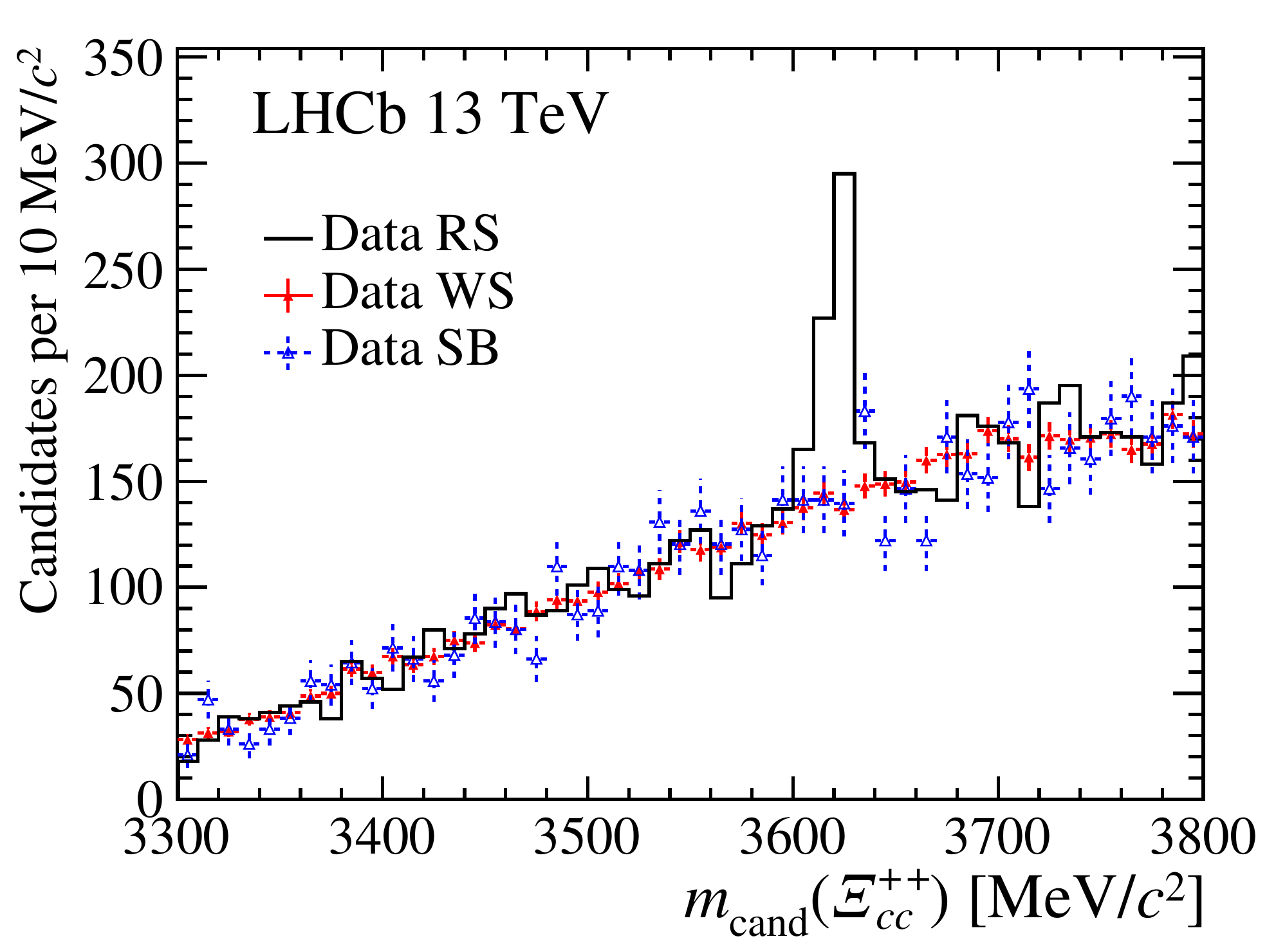}
  \end{center}
  \caption{
    Mass spectra of (\xiccfirstplot) \Lcp and (\xiccsecondplot) \Xiccpp candidates.
    The full selection is applied, except for the \Lcp mass requirement in the case of the \xiccfirstplot\ plot.
    For the $\Lcp$ mass distribution the (cross-hatched) signal and (vertical line) sideband regions are indicated;
    to avoid duplication, the histogram is filled only once in events that contain more than one \Xiccpp candidate.
    In the \xiccsecondplot\ plot the right-sign (RS) signal sample $\Xiccpp \to \Lcp \Km \pip \pip$ is shown, 
    along with the control samples: 
    \Lc sideband (SB) $\Lcp \Km \pip \pip$ candidates
    and wrong-sign (WS) $\Lcp \Km \pip \pim$ candidates,
    normalized to have the same area as the RS sample in the $\mcand(\Xiccpp)$ sidebands.
  }
  \label{fig:massWide}
\end{figure}

\begin{figure}
  \begin{center}
    \includegraphics[width=\xiccfeynwidth]{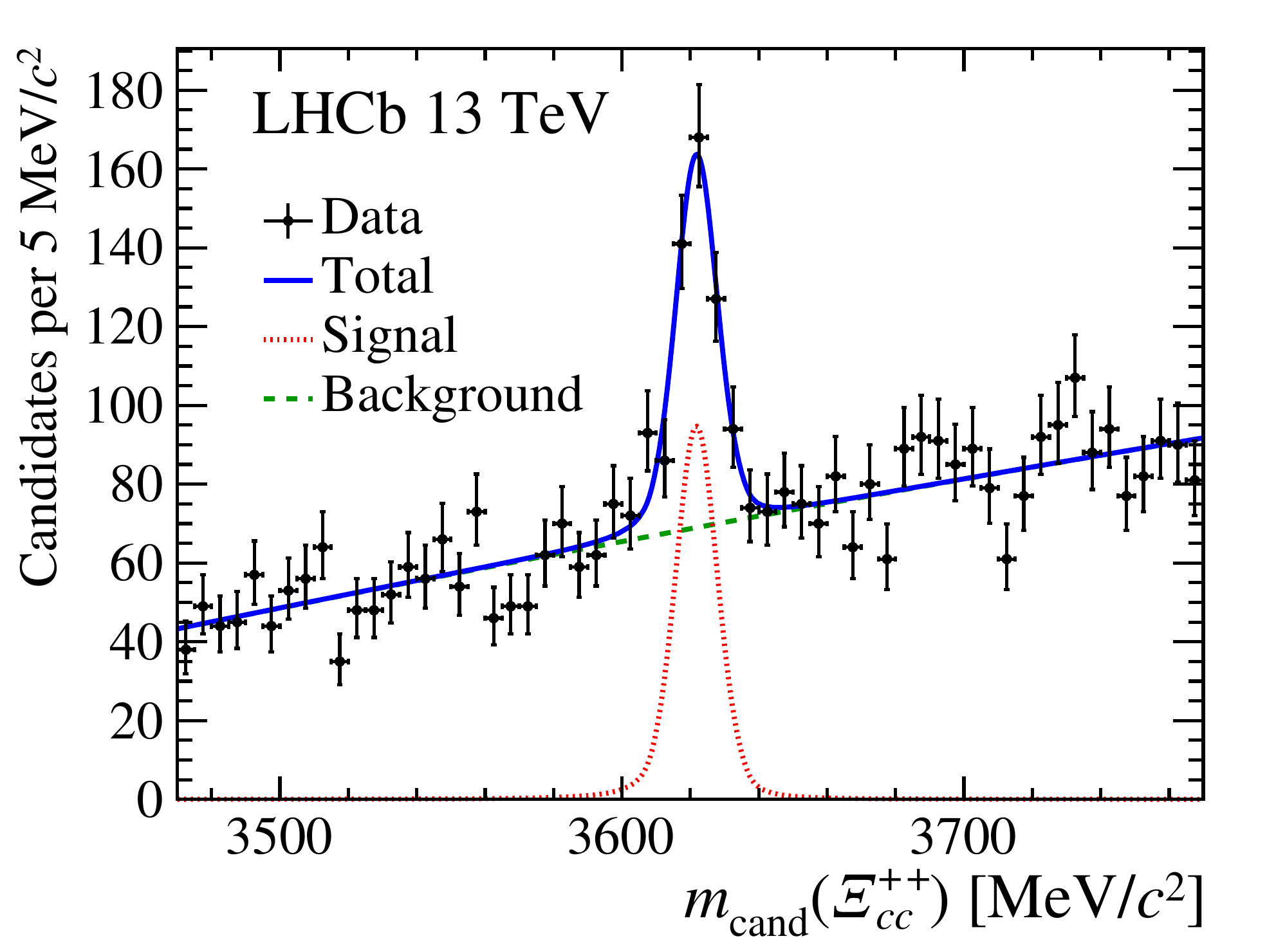}
  \end{center}
  \caption{
    Invariant mass
    distribution of $\Lcp\Km\pip\pip$ candidates with fit projections overlaid.
  }
  \label{fig:massFit}
\end{figure}

The selection described above
is then applied to data in the search region.
Figure~\ref{fig:massWide} shows the $\Lcp$ mass distribution, 
and the \Xiccpp mass spectra for candidates in the mass range $2270 < \mcand(\Lcp) < 2306$\mevcc.
A structure is visible in the signal mode at a mass of approximately 3620\mevcc. 
No significant structure is visible in the WS control sample,
nor for events in the \Lcp mass sidebands.
To measure the properties of the structure, an unbinned extended maximum likelihood fit is performed
to the invariant mass distribution in the restricted $\Lcp\Km\pip\pip$ mass window of $3620 \pm 150$\mevcc (Fig.~\ref{fig:massFit}).
The peaking structure is empirically described by a Gaussian function plus
a modified Gaussian function with power-law tails on both sides~\cite{Skwarnicki:1986xj}.
All peak parameters are fixed to values obtained from simulation
apart from the mass, yield,
and an overall resolution parameter.
The background is described by a second-order polynomial with parameters free to float in the fit.
The signal yield is measured to be
  $313 \pm 33$, corresponding to a local statistical significance in excess of $12\sigma$ when evaluated with a likelihood ratio test.
The fitted resolution parameter is
  $6.6 \pm 0.8$\mevcc, 
consistent with simulation. 
The same structure is also observed in the $\Lcp\Km\pip\pip$ spectrum in a $pp$ data sample collected by \lhcb at
$\sqrt{s}=8\tev$
(see supplemental material in Appendix~\ref{sec:SupplementalForPRL} for results from the 8\tev cross-check sample). 
%
The local statistical significance of the peak in the 8\tev sample is above $7\sigma$, and its mass
is consistent with that in the $13\tev$ data sample.

Additional cross-checks are performed confirming the robustness of the
observation.  The significance of the structure in the $\Lcp \Km \pip \pip$
final state remains above $12\sigma$ when fixing the resolution parameter in
the invariant mass fit to the value obtained from simulation, changing the
threshold value for the multivariate selector,
removing events containing multiple candidates in the fitted mass range,
or using an alternative
selection without a multivariate classifier. The significance also remains
above $12\sigma$ in a subsample of candidates for which the reconstructed
decay time exceeds five times its uncertainty. This is consistent with a
weakly decaying state and inconsistent with the strong decay of a resonance.
No fake peaking structures are observed in the control samples when requiring
various intermediate resonances to be present (\rhoz, \Kstarz, \Sigmacz,
\Sigmacpp, $\PLambda_c^{*+}$) nor are they observed when combining
\Xiccpp and \Lcp decay products.  The contributions of misidentified $\Ds \to
\Kp \Km \pip$ and $\Dp \to \Km \pip \pip$ decays are found to be negligible.

The sources of systematic uncertainty affecting the measurement of the $\Xiccpp$ mass (Table~\ref{tab:MassMeasurementSys}) 
include
the momentum-scale calibration, 
the event selection, 
the unknown $\Xiccpp$ lifetime,
the invariant mass fit model,
and the uncertainty on the \Lcp mass.
The momentum scale is calibrated with samples of $\jpsi\to\mumu$ and
$\Bp\to\jpsi\Kp$ decays~\cite{LHCb-PAPER-2011-035,LHCb-PAPER-2013-011}.
After calibration, an uncertainty 
of $\pm 0.03\%$ is assigned,
which corresponds to a systematic uncertainty of $0.22\mevcc$ on the reconstructed \Xiccpp mass.
The selection procedure is more efficient for vertices that are well separated from the PV,
and therefore preferentially retains longer-lived \Xiccpp candidates.
Because of a correlation between the reconstructed decay time and the reconstructed mass,
this induces a positive bias on the mass for both $\Xiccpp$ and $\Lcp$ candidates.
The effect is studied with simulation and the bias on the \Xiccpp mass is determined to be $+0.45 \pm 0.14$\mevcc
(assuming a lifetime of 333\fs), where the uncertainty is due to the limited size of the simulation sample.
A corresponding correction is applied to the fitted value in data.
To validate this procedure, the \Lcp mass in an inclusive sample is measured and corrected in the same way;
after the correction, the \Lcp mass is found to agree with the known value~\cite{PDG2017}.
The bias on the \Xiccpp mass depends on the unknown \Xiccpp lifetime,
introducing a further source of uncertainty on the correction.
This is estimated by repeating the procedure for other \Xiccpp lifetime hypotheses
between 200 and 700\fs. The largest deviation in the correction, $0.06$\mevcc,
is taken as an additional systematic uncertainty.
Final-state photon radiation also causes a bias in the measured mass,
which is determined to be $-0.05$\mevcc with simulation~\cite{Golonka:2005pn}.
The uncertainty on this correction is approximately $0.01$\mevcc
and is neglected.
The dependence of the measurement on the fit model is estimated by
varying the shape parameters that are fixed according to simulation,
by using alternative signal and background models,
and by repeating the fits in different mass ranges.
The largest deviation seen in the mass, $0.07$\mevcc,
is assigned as a systematic uncertainty.
Finally, since the \Xiccpp mass is measured relative to the \Lcp mass,
the uncertainty of $0.14$\mevcc on the world-average value of the latter is included.
After taking these systematic effects into account
and combining their uncertainties (except that on the \Lcp mass) in quadrature,
the \Xiccpp mass is measured to be
\XiccmassWithLcError.
The mass difference between the \Xiccpp and \Lcp states is
\XiccDeltaMass.

\begin{table}
    \caption{Systematic uncertainties on the \Xiccpp mass measurement. }
\centering
\begin{tabular}{lc}
\hline
Source                       & Value $[\mevcc]$\\
\hline
Momentum-scale calibration   & 0.22\\
Selection bias correction    & 0.14\\
Unknown \Xiccpp lifetime     & 0.06\\
Mass fit model               & 0.07\\
\hline
Sum of above in quadrature            & 0.27\\
\hline
$\Lc$ mass uncertainty       & 0.14\\
\hline
\end{tabular}\label{tab:MassMeasurementSys}
\end{table}

In summary, a highly significant structure is observed in the final state $\Lcp \Km \pip \pip$ 
in a $pp$ data sample collected by \lhcb at $\sqrt{s}=13\tev$,
with a signal yield of $313\pm 33$.
The mass of the structure is measured to be 
\XiccmassWithLcError,
where the last uncertainty is due to the limited knowledge of the \Lcp mass,
and its width is consistent with experimental resolution. 
The structure is confirmed with consistent mass in
a data set collected by LHCb at $\sqrt{s}=8\tev$.
The signal candidates have significant decay lengths, and
the signal remains highly significant
after a minimum lifetime requirement of
approximately five times the expected decay-time resolution is imposed.
This state is therefore
incompatible with a strongly decaying particle but is consistent with the expectations for the weakly decaying \Xiccpp baryon.
The mass of the observed \Xiccpp state is greater
than that of the \Xiccp peaks reported by the SELEX collaboration~\cite{Mattson:2002vu,Ocherashvili:2004hi}
by $103 \pm 2$\mevcc.
This difference would imply an isospin splitting vastly larger than that seen
in any other baryon system and is inconsistent with the expected size
of a few~\mevcc~\cite{Hwang:2008dj,Brodsky:2011zs,Karliner:2017gml}.
Consequently, while the state reported here is consistent with
most theoretical expectations for the \Xiccpp baryon, it is inconsistent
with being an
isospin partner to the \Xiccp state reported previously by the SELEX
collaboration.

\section*{Acknowledgements}

\noindent We thank Chao-Hsi Chang, Cai-Dian L\"u, Xing-Gang Wu, and Fu-Sheng Yu for frequent and interesting discussions on the production and decays of double-heavy-flavor baryons.
We express our gratitude to our colleagues in the CERN
accelerator departments for the excellent performance of the LHC. We
thank the technical and administrative staff at the LHCb
institutes. We acknowledge support from CERN and from the national
agencies: CAPES, CNPq, FAPERJ and FINEP (Brazil); MOST and NSFC (China);
CNRS/IN2P3 (France); BMBF, DFG and MPG (Germany); INFN (Italy); 
NWO (The Netherlands); MNiSW and NCN (Poland); MEN/IFA (Romania); 
MinES and FASO (Russia); MinECo (Spain); SNSF and SER (Switzerland); 
NASU (Ukraine); STFC (United Kingdom); NSF (USA).
We acknowledge the computing resources that are provided by CERN, IN2P3 (France), KIT and DESY (Germany), INFN (Italy), SURF (The Netherlands), PIC (Spain), GridPP (United Kingdom), RRCKI and Yandex LLC (Russia), CSCS (Switzerland), IFIN-HH (Romania), CBPF (Brazil), PL-GRID (Poland) and OSC (USA). We are indebted to the communities behind the multiple open 
source software packages on which we depend.
Individual groups or members have received support from AvH Foundation (Germany),
EPLANET, Marie Sk\l{}odowska-Curie Actions and ERC (European Union), 
Conseil G\'{e}n\'{e}ral de Haute-Savoie, Labex ENIGMASS and OCEVU, 
R\'{e}gion Auvergne (France), RFBR and Yandex LLC (Russia), GVA, XuntaGal and GENCAT (Spain), Herchel Smith Fund, The Royal Society, Royal Commission for the Exhibition of 1851 and the Leverhulme Trust (United Kingdom).

\appendix
\clearpage

\section{Appendix: Supplemental material}
%
\label{sec:SupplementalForPRL}

The Letter describes the observation of a narrow structure in the
$\Lcp \Km \pip \pip$ mass spectrum in a sample of data collected
by the LHCb experiment in 2016 at a center-of-mass energy of 13\tev,
corresponding to an integrated luminosity of $1.7$\invfb.
In addition, as a cross-check, a similar study is carried out
on a separate data sample collected in 2012 at a center-of-mass energy
of 8\tev, corresponding to an integrated luminosity of $2.0$\invfb.
The 13\tev sample has greater sensitivity, due both to
an increase in the expected cross-section at higher center-of-mass energy
and to improvements in the online selection between the
data-taking periods. Nonetheless, a smaller but still highly
significant signal is also found in the 8\tev sample, with
properties fully compatible with those of the signal seen
in the 13\tev sample. This serves as a useful, and statistically
independent, validation.
In this supplemental material,
the differences between the two data samples are outlined and
results from the cross-check 8\tev sample are shown.

Data taken during 2012 follow an event processing model in
which events are first required to pass a multi-level online event
selection. The online selection used for this study is the same
as that described in Ref.~\cite{LHCb-PAPER-2013-049}.
The events are then analyzed offline and the decay chain
\XiccppDecay is reconstructed following the procedure described
in the Letter. The $\Xiccpp$ candidates are required to
pass the same series of selection criteria as for the 13\tev
sample, as well as three additional requirements
(on the \pt of the products of the \Lc decay,
 on the particle identification information of the \pip from the \Lc decay,
 and on the distances of closest approach of the decay products of the \Xiccpp to one another)
that were applied as part of an initial event filtering pass.
Candidates are also required to pass the multivariate selector described in the Letter.
For consistency, the same selector used in the 13\tev sample was applied to the 8\tev sample.
However, the threshold on the selector output was reoptimized
with control samples with a center-of-mass energy of 8\tev.

\begin{figure}
  \begin{center}
     \includegraphics[width=\xiccplotwidth]{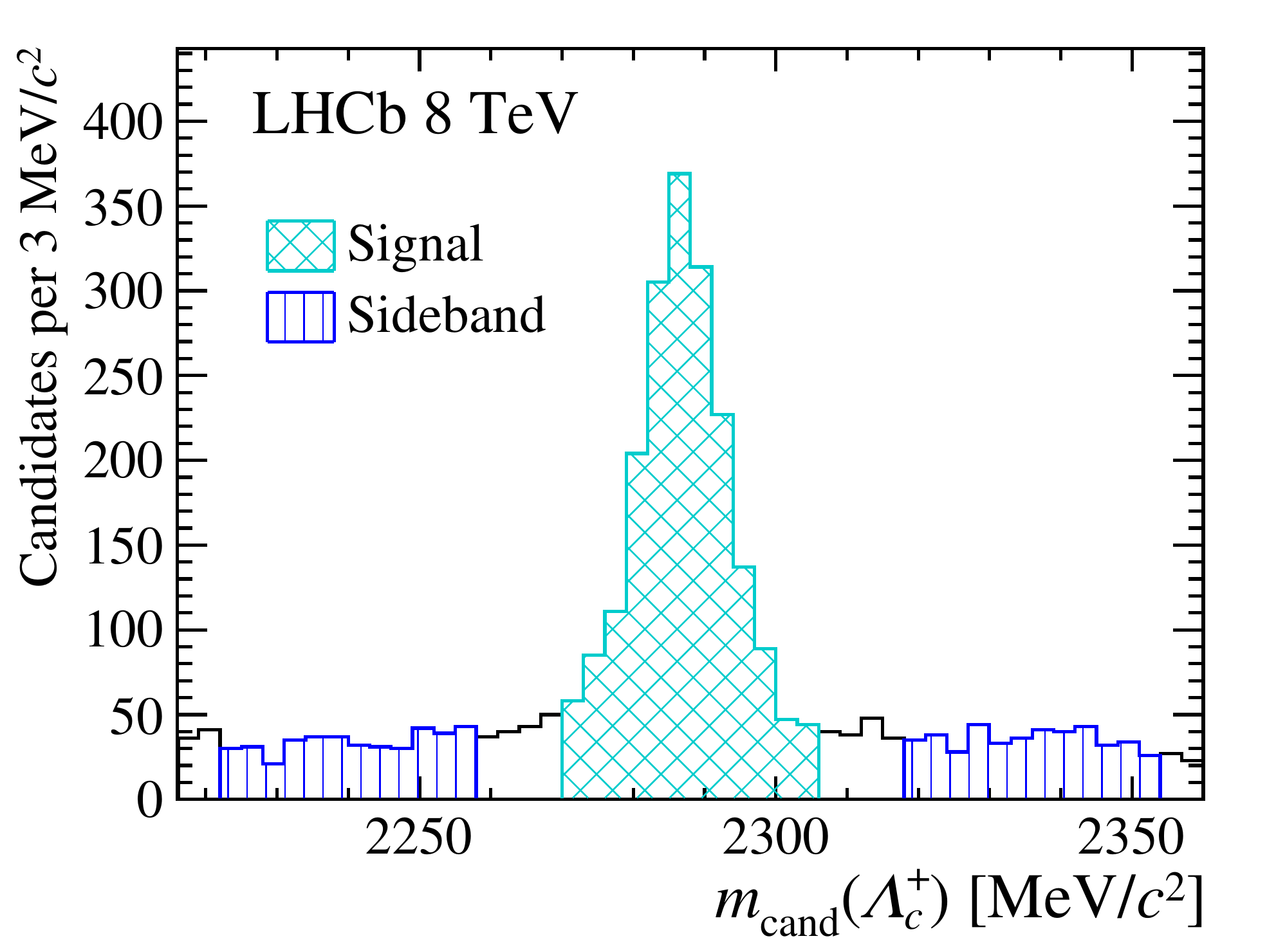}
     \includegraphics[width=\xiccplotwidth]{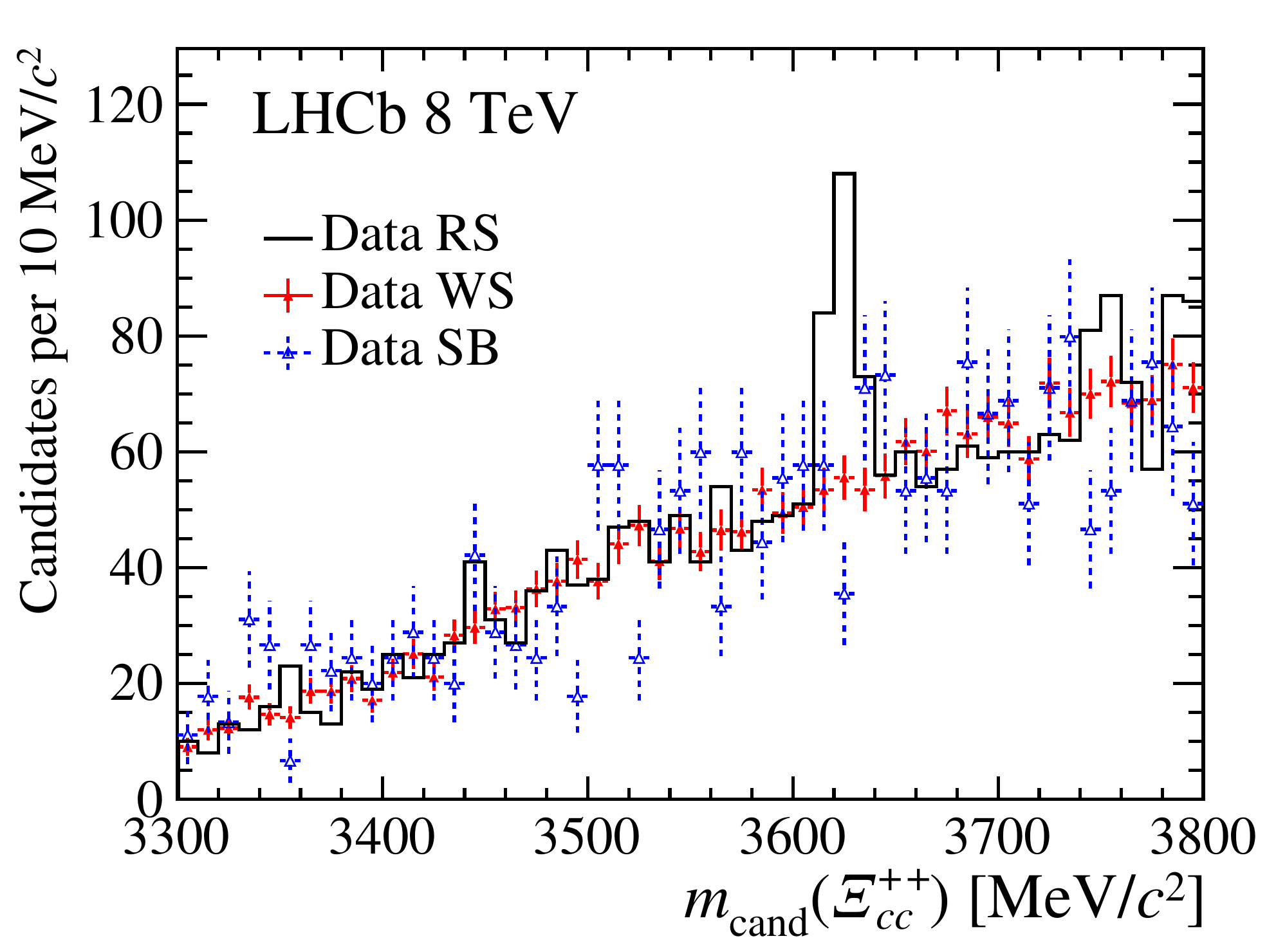}
   \end{center}
   \caption{
     Mass spectra of (\xiccfirstplot) \Lcp and (\xiccsecondplot) \Xiccpp candidates in the $8\protect\tev$ data sample.  
     The full selection is applied, except for the \Lcp mass requirement in the case of the \xiccfirstplot\ plot.
     For the $\Lcp$ mass distribution the (cross-hatched) signal and (vertical lines) sideband regions are indicated;
     to avoid duplication, the histogram is filled only once in events that contain more than one \Xiccpp candidate.
     In the \xiccsecondplot\ plot the right-sign (RS) signal sample $\Xiccpp \to \Lcp \Km \pip \pip$ is shown, 
     along with the control samples: 
     \Lc sideband (SB) $\Lcp \Km \pip \pip$ candidates
     and wrong-sign (WS) $\Lcp \Km \pip \pim$ candidates,
     normalized to have the same area as the RS sample in the $\mcand(\Xiccpp)$ sidebands.
   }
   \label{fig:SupplementalForPRL:massWide}
\end{figure}

\begin{figure}
  \begin{center}
    \includegraphics[width=\xiccfeynwidth]{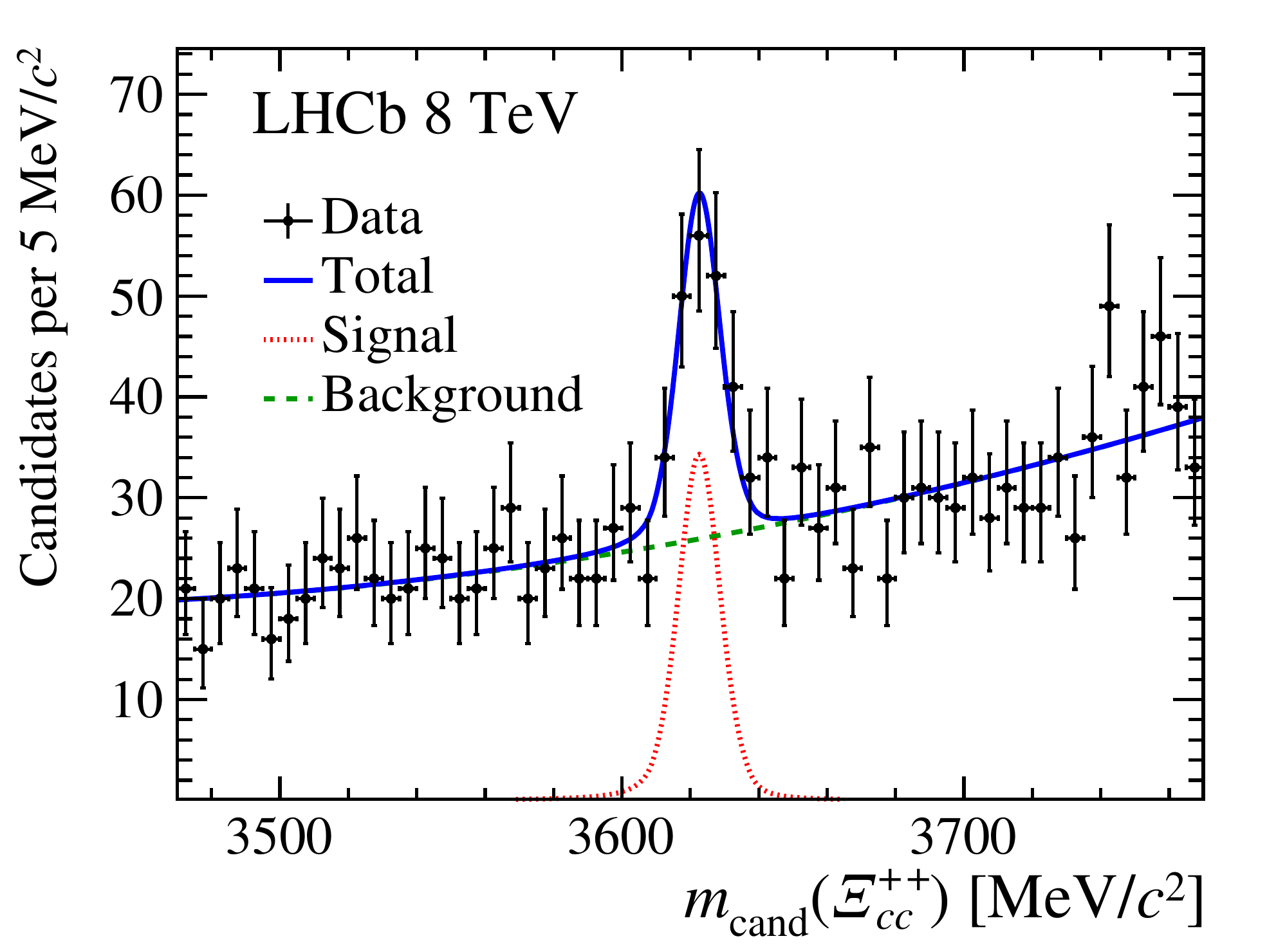}
  \end{center}
  \caption{
    Invariant mass distribution of $\Lcp\Km\pip\pip$ candidates for the 8\,TeV  
   data sample with fit
   projections overlaid.
  }
  \label{fig:SupplementalForPRL:massFit}
\end{figure}

Figure~\ref{fig:SupplementalForPRL:massWide} shows the \Lcp and \Xiccpp mass spectra in the 8\tev sample after the final selection. 
As with the 13\tev sample, a narrow structure is visible in the
signal mode but no structure is seen in the control samples.
The fit procedure described in the Letter is applied to the 8\tev right-sign sample, 
and the results are shown in Fig.~\ref{fig:SupplementalForPRL:massFit}. 
The signal yield is measured to be
$113 \pm 21$, and corresponds to a statistical significance in excess
of seven standard deviations. The fitted mass differs from that in the 13\tev
sample by $0.8 \pm 1.4$\mevcc (where the uncertainty is statistical only).
The fitted resolution parameter is $6.6 \pm 1.4$\mevcc, consistent with that in the 13\tev sample
and with the value expected from simulation. The resolution parameter is the weighted average of the widths of the two Gaussian functions of the signal mass fit model. 
Thus, the fitted properties of the
structures seen in the two samples are consistent, and we conclude that they
are associated with the same physical process.
Combined with the yield of $313 \pm 33$ in the 13\tev data sample,
the total signal yield in the two samples is $426 \pm 39$.

\clearpage

\addcontentsline{toc}{section}{References}
\setboolean{inbibliography}{true}
\bibliographystyle{LHCb}
\bibliography{main,LHCb-PAPER,LHCb-CONF,LHCb-DP,LHCb-TDR,xicc-refs}

\newpage

\newpage
\centerline{\large\bf LHCb collaboration}
\begin{flushleft}
\small
R.~Aaij$^{40}$,
B.~Adeva$^{39}$,
M.~Adinolfi$^{48}$,
Z.~Ajaltouni$^{5}$,
S.~Akar$^{59}$,
J.~Albrecht$^{10}$,
F.~Alessio$^{40}$,
M.~Alexander$^{53}$,
A.~Alfonso~Albero$^{38}$,
S.~Ali$^{43}$
% ...

{\footnotesize \it
% ...
$ ^{5}$Clermont Universit{\'e}, Universit{\'e} Blaise Pascal, CNRS/IN2P3, LPC, Clermont-Ferrand, France\\
% ...
$ ^{10}$Fakult{\"a}t Physik, Technische Universit{\"a}t Dortmund, Dortmund, Germany\\
% ...
$ ^{38}$ICCUB, Universitat de Barcelona, Barcelona, Spain\\
$ ^{39}$Universidad de Santiago de Compostela, Santiago de Compostela, Spain\\
$ ^{40}$European Organization for Nuclear Research (CERN), Geneva, Switzerland\\
% ...
$ ^{43}$Nikhef National Institute for Subatomic Physics, Amsterdam, The Netherlands\\
% ...
$ ^{48}$H.H. Wills Physics Laboratory, University of Bristol, Bristol, United Kingdom\\
% ...
$ ^{53}$School of Physics and Astronomy, University of Glasgow, Glasgow, United Kingdom\\
% ...
$ ^{59}$University of Cincinnati, Cincinnati, OH, United States\\
% ...
}
\end{flushleft}

\end{document}